\documentclass[twoside]{article}

\usepackage{lipsum} % Package to generate dummy text throughout this template
\usepackage{aas_macros}

\usepackage[sc]{mathpazo} % Use the Palatino font
\usepackage[T1]{fontenc} % Use 8-bit encoding that has 256 glyphs
\usepackage{microtype} % Slightly tweak font spacing for aesthetics

\usepackage[hmarginratio=1:1,top=32mm,columnsep=20pt]{geometry} % Document margins
\usepackage{multicol} % Used for the two-column layout of the document
\usepackage[hang, small,labelfont=bf,up,textfont=it,up]{caption} % Custom captions under/above floats in tables or figures
\usepackage{booktabs} % Horizontal rules in tables
\usepackage{float} % Required for tables and figures in the multi-column environment - they need to be placed in specific locations with the [H] (e.g. \begin{table}[H])
\usepackage{hyperref} % For hyperlinks in the PDF
\usepackage{natbib} %- my add
\usepackage{graphicx}
\usepackage{caption} %subfigures - my add
\usepackage{subcaption} %subfigures - my add
\usepackage{lettrine} % The lettrine is the first enlarged letter at the beginning of the text
\usepackage{paralist} % Used for the compactitem environment which makes bullet points with less space between them
\usepackage{wasysym}
\usepackage{stmaryrd}

\usepackage{abstract} % Allows abstract customization
\usepackage{color}

\usepackage{ulem}

\usepackage{titlesec} % Allows customization of titles
\renewcommand\thesection{\Roman{section}} % Roman numerals for the sections
\renewcommand\thesubsection{\thesection.\arabic{subsection}} % Roman numeralsfor subsections
\titleformat{\section}[block]{\large\scshape\centering}{\thesection.}{1em}{} % Change the look of the section titles
\titleformat{\subsection}[block]{\large}{\thesubsection.}{1em}{} % Change the look of the section titles

\usepackage{fancyhdr} % Headers and footers
\title{\vspace{-15mm}\fontsize{24pt}{10pt}\selectfont\textbf{Making It Rain: How Giving Me Telescope Time Can Reduce Drought}} % Article title

\author{
\large
\textsc{Michael B. Lund$^1$}\\%, Robert J. Siverd$^2$, and Ponder Stibbons$^3$}\\%\thanks{A thank you or further information}\\[2mm] % Your name
\normalsize $^1$CalTech/IPAC-NExScI\\
\normalsize \href{mailto:mlund@ipac.caltech.edu}{mlund@ipac.caltech.edu} % Your email address
\vspace{-5mm}
}
\date{}

\begin{document}

\maketitle % Insert title

\thispagestyle{fancy} % All pages have headers and footers

%----------------------------------------------------------------------------------------
%	ABSTRACT
%----------------------------------------------------------------------------------------

\begin{abstract}

\noindent In this paper we assess the correlation between recent observing runs (2018 and 2019) and inclement weather, and demonstrate that these observing runs have seen much more rainfall than would otherwise be expected, an increase of over 200\%. We further look at a number of observatory sites in areas that are facing or will face drought, and suggest that a strong environmental benefit would follow from telescope allocation committees providing us an inordinate amount of telescope time at facilities located around the globe.

\end{abstract}

%----------------------------------------------------------------------------------------
%	ARTICLE CONTENTS
%----------------------------------------------------------------------------------------

\begin{multicols}{2} % Two-column layout throughout the main article text

\section{Introduction}
\lettrine[nindent=0em,lines=3]{T}he concept of rainfall being directly linked to a single individual has a long history of speculation within the literature \citep{Kinney1954, Lovy1967, Braybrooks1996}. Indeed, even the possibility of harnessing this effect is discussed in \citet{Adams1984}.

The first demonstration in a laboratory setting of artificial precipitation was achieved by using silver iodine to stimulate the formation of ice crystals necessary for rain \citep{Vonnegut1949}. It was further shown that by intentionally seeding clouds with silver iodine, it is possible to create artificial ice crystal formation \citep{Vonnegut1952} a necessary prequesite for rain \citep{Nye1995}. This process has been demonstrated, and can be implemented to create human-triggered precipitation \citep{French2018}. Furthermore, it has  been argued that ionizing radiation can function as a seed in a similar fashion, removing the need for cloud doping with particle deposition \citep{Svensmark2017}.

Though it has been suggested that humans can artificially impact the weather, the broader question of the anthropogenic influence over climate is much more robust. Indeed, the impact of humans on the climate has received enough study to raise this question to the level of scientific fact \citep{IPCC2018}. This includes human activity resulting in increases in rain \citep{Allan2011}, and so it is not impossible to suggest that small-scale corrections in weather and climate can be made with targeted actions to increase rainfall.

In this paper, we use a combination of historical data and our own observations (presented in Section~\ref{Data}), and demonstrate in Section~\ref{Methods}) that our observing runs have had significantly more rainy days and rainfall than expected, even when accounting for the seasonal distribution of our observing runs, with estimated rainfall at roughly 350\% the expected amount. We make the case that the author being awarded additional telescope time at observatories located in areas prone to drought can improve deficiencies in rainfall (Section~\ref{Discussion}, and we summarize our results in Section~\ref{Summary}.

%------------------------------------------------
\section{Data} \label{Data}
We rely on two data sets for this work, both relating to observing runs that took place over 2018 and 2019 at Palomar Observatory on Mount Palomar in San Diego County, California. Palomar Observatory is owned and operated by the California Institute of Technology. The historic climate data is outlined in Subsection~\ref{data:hist} and our observing track record is discussed in Subsection~\ref{data:obs}.

\subsection{Mount Palomar Climate} \label{data:hist}
In order to characterize the baseline climate for Mount Palomar, we gathered rainfall data for Mount Palomar (ZIP 92060) from bestplaces.net, a site that sources a range of climate data with monthly resolution\footnote{\url{https://www.bestplaces.net/climate/zip-code/california/palomar_mountain/92060}}. Of particular interest to our inquiry is the number of rainy days in a month and the inches of rain per month, both displayed in Table~\ref{tab:Obs_run} (yearly totals may not match source due to rounding differences).
\begin{table}[H]
\caption{Mount Palomar Climate Data}
\label{tab:Climate_data}
\begin{tabular}{lll}
Month & Rainy Days & Inches of Rain \\
\hline
Jan   & 6.9        & 4.7            \\
Feb   & 7.9        & 5.0            \\
Mar   & 7.2        & 4.0            \\
Apr   & 4.7        & 1.7            \\
May   & 2.4        & 0.5            \\
Jun   & 0.6        & 0.1            \\
Jul   & 1.3        & 0.3            \\
Aug   & 1.8        & 0.5            \\
Sep   & 2.0        & 0.5            \\
Oct   & 3.2        & 1.0            \\
Nov   & 4.4        & 2.0            \\
Dec   & 6.2        & 3.4            \\
\hline
Total & 48.6       & 23.7          
\end{tabular}
\end{table}

\begin{figure*}[htb]
  \begin{center}
   \includegraphics[width=.45\textwidth]{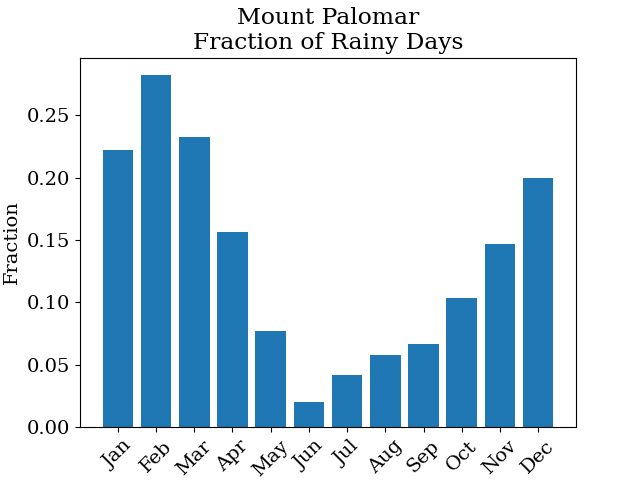}
   \includegraphics[width=.45\textwidth]{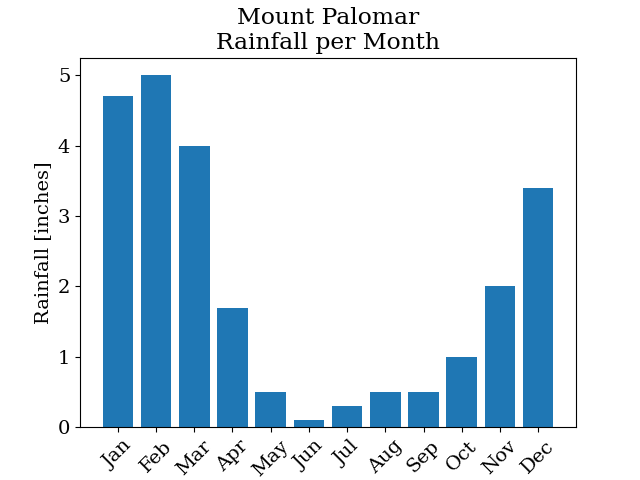}
   \includegraphics[width=.45\textwidth]{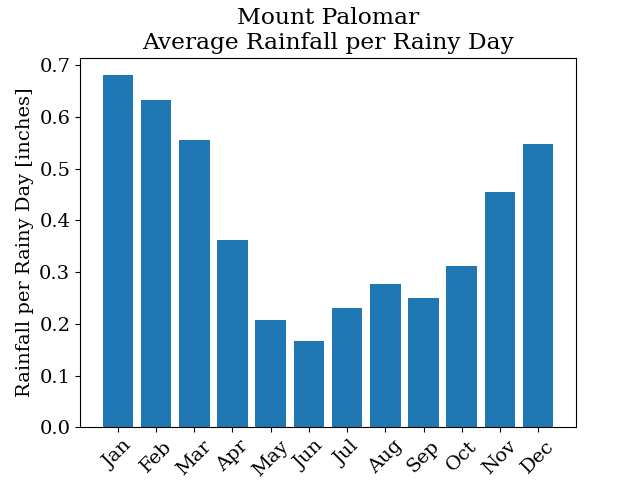}
  \end{center}
  \caption{Climate data for Mount Palomar. The top left figure shows the average number of rainy days per month. In the top right, we show the average monthly rainfall. In the bottom figure, we show the average rainfall per rainy day, based on the data in the other two figures.}
  \label{fig:climate}
\end{figure*}

This data is visualized in Figure~\ref{fig:climate}. We take the fraction of days in the month that were rainy and combine this with the total amount of rain per month to calculate the average rainfall that occurs on an individual rainy day.

Additional weather data is available from Weather Underground\footnote{\url{www.wunderground.com}}, that we refer to here but do not reproduce in this paper. We use this data as a supplemental approach to our main work, but note that while Weather Underground provides both historical and average rainfall with daily binning, the closest weather station to Palomar Mountain with historical data available through the site is the McClellan–Palomar Airport Weather Station in Carlsbad, CA, about 45 kilometers WSW of Palomar Mountain and 1600 meters lower. This distance and change in elevation results in significant uncertainty in applying this data to Mount Palomar. Individual days were queried from Weather Underground as detailed in Section~\ref{Methods}.

\subsection{Observing Runs} \label{data:obs}
Here we consider nine observing runs, all of which were scheduled on the 200-in Hale Telescope, and using an instrument which requires traveling to Palomar Observatory to observe. Travel actually occurred for 8 of these runs, however the Feb 21-22, 2019 run was significantly impacted by weather such that the top of the mountain could not be safely reached.  We include this case in our analysis because the key element is the intention to go to Palomar Observatory, not necessarily the action of it, as it has been argued that the brain treats activates in the same fashion for both thinking about and carrying out an action \citep{Ptak2017}. We also do not include any observing runs that took place that the author was not scheduled to travel for. The full list of observing runs and the number of rainy nights are listed in Table~\ref{tab:Obs_run}, with the dates corresponding to the start of each night (i.e. Feb 21-22, 2019 was the nights of Feb 21 and 22, but did not end until the morning of Feb 23).
\begin{table}[H]
\caption{Weather on Observing Runs}
\label{tab:Obs_run}
\begin{tabular}{ll}
Observing Run   & Rainy Nights \\
\hline
Feb 21-22, 2019 & 2          \\
Mar 20-21, 2019 & 2          \\
Apr 17, 2019    & 0          \\
May 15-16, 2019 & 2          \\
Jun 11-13, 2019 & 3          \\
Jun 22, 2019    & 1          \\
Jul 22, 2019    & 1          \\
Nov 8-9, 2019   & 2          \\
Mar 9-10, 2020  & 2         
\end{tabular}
\end{table}
We had 8 nights where rain or snow prevented observations. Data taken on the successful nights has all been reduced and submitted to ExoFOP-TESS\footnote{\url{https://exofop.ipac.caltech.edu/tess/}}, and a more in-depth analysis of those observations will appear in Lund et al (in prep).

\section{Methods} \label{Methods}
\begin{figure*}[!htb]
  \begin{center}
   \includegraphics[width=.45\textwidth]{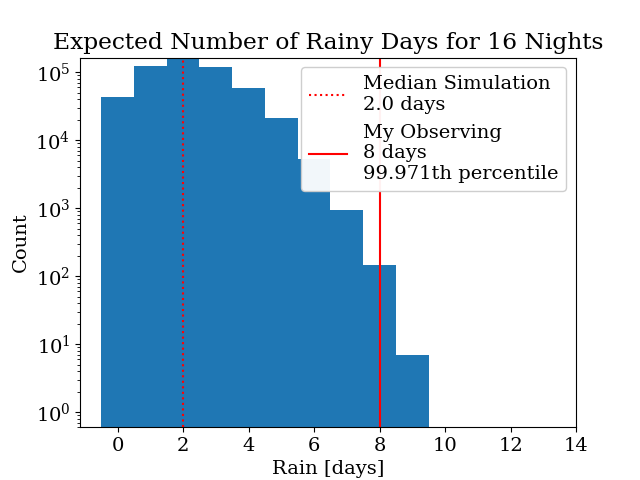}
   \includegraphics[width=.45\textwidth]{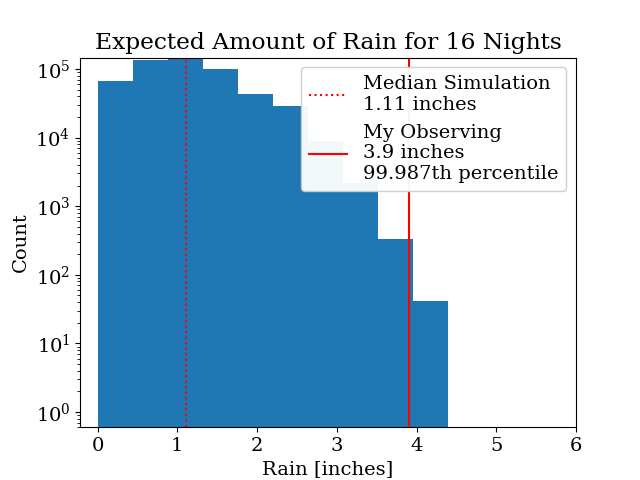}
  \end{center}
  \caption{We simulated 525,600 sets of comparable observing runs. On the left we show the distribution of how many rainy nights to expect, with the median being one fourth of what we observed. On the right we show the expected rainfall is much less than our observed rainfall. In both cases, our observed rainfall exceeds the 99th percentile.}
  \label{fig:sims}
\end{figure*}

In order to compare our observing run to what could be otherwise expected for rainfall, we created 525,600 simulated observing runs. For each run, we preserve the months that we observed in, and for each night determine if the day is rainy based on the fraction of rainy days in a month (Figure~\ref{fig:climate}, top left). If the day is rainy, then we assign to that day the average amount of rain per rainy day in that month (Figure~\ref{fig:climate}, bottom). The distributions of total number of rainy days and total amount of rain expected are shown in Figure~\ref{fig:sims}. We determine our own approximated rainfall by summing the average rainy day rainfall for all nights where we suffered inclement weather. We find that the rainfall we observed exceeds the $99^{th}$ percentile, with 8 days of rain being four times the median of 2 days of rain expected, and 3.9 inches of rain being about 350\% of the expected rainfall. Indeed, in 16 nights over 14 months (February 2019 to March 2020), our observing runs account for just under 4\% of nights, but 13\% of rainy days and 12\% of expected rainfall.

As an alternative approach, we also use the Weather Underground data previously discussed in Section~\ref{data:hist}. This data source provides daily rain totals, but at the somewhat distant location of Carlsbad, CA. As the Weather Underground data defines the change of day at midnight, we also include the day that our observing concludes on in our queries. Summaries for each observing run are listed in Table~\ref{tab:Wunderground}, with the total rainfall of our observing runs and the total average rainfall on the same dates included at the bottom. While we note the greater uncertainty in this measurement, we still find an increase in rainfall using this method of about 60\%, which is consistent with a significant increase in rainfall during our observing runs.

\begin{table}[H]
\caption{Weather Underground Data for McClellan–Palomar Airport Weather Station}
\label{tab:Wunderground}
\begin{tabular}{lll}
Observing Run   & \multicolumn{1}{c}{\begin{tabular}[c]{@{}c@{}}Total\\ Rainfall\\ {[}inches{]}\end{tabular}} & \multicolumn{1}{c}{\begin{tabular}[c]{@{}c@{}}Average\\ Rainfall\\ {[}inches{]}\end{tabular}} \\
\hline
Feb 21-22, 2019 & 0.14                                                                                        & 0.37                                                                                      \\
Mar 20-21, 2019 & 0.35                                                                                        & 0.17                                                                                      \\
Apr 17, 2019    & 0.00                                                                                        & 0.06                                                                                      \\
May 15-16, 2019 & 0.18                                                                                        & 0.01                                                                                      \\
Jun 11-13, 2019 & 0.00                                                                                        & 0.02                                                                                      \\
Jun 22, 2019    & 0.03                                                                                        & 0.00                                                                                      \\
Jul 22, 2019    & 0.00                                                                                        & 0.00                                                                                      \\
Nov 8-9, 2019   & 0.00                                                                                        & 0.09                                                                                      \\
Mar 9-10, 2020  & 0.76                                                                                        & 0.17                                                                                      \\
\hline
Total           & 1.46                                                                                        & 0.89                                                                                     
\end{tabular}
\end{table}

\section{Discussion}\label{Discussion}
\begin{figure*}[!htb]
  \begin{center}
   \includegraphics[width=\textwidth]{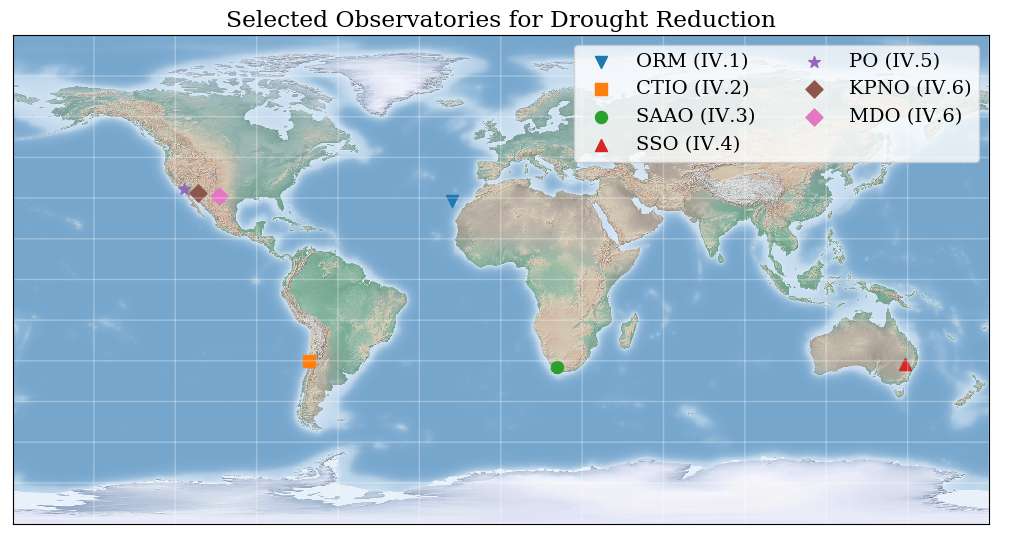}
  \end{center}
  \caption{All observatories that we have selected as ideal for getting telescope time to counter drought, spanning 5 continents. We label each site with the observatory abbreviation and the subsection that they are discussed in.}
  \label{fig:map}
\end{figure*}

Most modern observatories are located in climates favorable to observing. Two of the most important site qualities required to increase on-sky time and improve seeing are low cloud cover and low precipitable water vapor \citep{Aksaker2020}. However, sites that regularly exhibit both of these qualities are more susceptible to drought conditions. Given that we have already demonstrated that our observing runs result an increase of rainfall, we now discuss where telescope time can be allocated to us such that it can provide maximal benefits to the surrounding environment. We address several regions, and show the locations of the observatories in Figure~\ref{fig:map}.

\subsection{Canary Islands}
The Canary Islands of Spain have seen several drought seasons in the last ten years, and these patterns pose a particular risk for some of the niche ecosystems that exist on the islands \citep{Olano2017}. This region may require the stabilization provided by greater rainfall to counter the damage from global warming in the coming years. The island of La Palma also is home to Roque de los Muchachos Observatory (ORM), frequently cited as the second-best location for optical astronomy in the northern hemisphere, and features telescopes like the 10.4 m Gran Telescopio Canarias.

\subsection{Chile}
Central Chile has experienced what has been referred to as a 'mega-drought' after facing a series of dry years since 2010, with rainfall down 20-40\% from normal \citep{Garreaud2020}. This region is home to the majority of Chile's population, supporting approximately 10 million people. Additional rainfall would prevent the further loss of livestock and replenish reservoirs that are running low. On the northern boundary of this region are several astronomical facilities: the Cerro Tololo Inter-American Observatory (CTIO) includes the 4m Victor M. Blanco Telescope on Cerro Tololo and the 4.1m Southern Astrophysical Research Telescope on nearby Cerro Panch\'{o}n.

\subsection{South Africa}
In 2018, Cape Town, South Africa was at a very real risk of being the first major city to run out of water due to a long-term trend in reduced winter rainfall and significant seasonal variability in rainfall \citep{Burls2019}. Additional precipitation would be a vital contribution to Cape Town's water supply, and so the rains are very important \citep{Toto1982}. Approximately 250-300 km northeast of Cape Town is Sutherland and the South African Astronomical Observatory (SAAO). As one of the largest observatories in Africa, the site is also home to the $\sim$10m Southern African Large Telescope.

\subsection{Southeast Australia}  
The wheat belt that stretches across southeast Australia (predominantly in New South Wales) is expected to see an increasing frequency of droughts, and these droughts will have greater severity \citep{Feng2019}. More critically, this is a region of key agricultural output, and further decreases in rainfall threaten to put increased stress on the regional food supply. Located roughly within this region of the sunburnt country \citep{Mackellar1908} is Siding Spring Observatory (SSO), outside of Coonabarabran. Siding Spring Observatory is the site of numerous telescopes, the largest being the 3.9m Anglo-Australian Telescope.

\subsection{Southern California}
A study of the San Diego region showed that rainfall has declined over the timespan from 1985-2017 \citep{Mosase2019}, and that droughts can play a key role in devastating fires in Southern California \citep{Taylor1970}. Palomar Observatory (PO) is located about 100 km NNE of San Diego and falls within the San Diego region. While Southern California is known for having very little rain, when it does rain the rain can be quite heavy \citep{Hammond1972}. We have already demonstrated the utility of additional observing time at Palomar Observatory to increase rainfall.

\subsection{Southwest United States}
Throughout the Southwest of the United States (Arizona, New Mexico, Texas, and Oklahoma) there has been a major decrease in soil water storage. Climate change has caused decreases in atmospheric water input, which has been a major to dominant factor in this change in soil water storage \citep{Liu2019}. This region also includes major population centers that now are at risk of running out of water, such as Phoenix, Arizona and El Paso, Texas, and could greatly benefit from additional rain\footnote{\url{https://weather.com/forecast-change/news/2019-06-03-5-us-cities-that-could-potentially-run-out-of-water}}. The southwest United States is also rich in observing facilities. Kitt Peak National Observatory (KPNO) is located in southern Arizona, with almost two dozen telescopes, including the Mayall 4-meter Telescope. Over 700 km east in this same region is McDonald Observatory (MDO), with 4 research telescopes, including the 10m Hobby-Eberly Telescope. Prior unpublished observations have also confirmed that the author's presence at McDonald Observatory correlated with higher levels of rainfall than would be otherwise expected for the time of year that the observing runs took place.

\section{Summary}\label{Summary}
In this paper we have demonstrated how our observing runs at Palomar Observatory have corresponded with an increase in rainy nights. Our 16 nights of observing saw $\sim$4 inches of rain, more than three times the median expected rainfall of 1.1 inches. The chance of this much rain by chance is significantly less than 1\%. We have further discussed that many premier facilities globally are located in areas that are prone to drought, spanning five continents. We believe that this provides a valuable opportunity for drought conditions to be reduced by awarding the author significant telescope time.

\section{Acknowledgements}
The author thanks Savannah R. Jacklin for her valuable feedback on this manuscript. The author also thanks David R. Ciardi for not firing him after three consecutive unsuccessful observing runs, and the Palomar staff for ensuring high quality observations when it was not raining.

This research made use of Astropy,\footnote{http://www.astropy.org} a community-developed core Python package for Astronomy \citep{astropy:2013, astropy:2018}.
%----------------------------------------------------------------------------------------
%	REFERENCE LIST
%----------------------------------------------------------------------------------------

\bibliographystyle{apalike}
\bibliography{main}

%----------------------------------------------------------------------------------------

\end{multicols}

\end{document}